# HIGHER ORDER MODE DAMPER FOR LOW ENERGY RHIC ELECTRON COOLER SRF BOOSTER CAVITY

Binping Xiao†, A. Fedotov, H. Hahn, D. Holmes, G. McIntyre, C. Pai, S. Seberg, K. Smith, R. Than, P. Thieberger, J. Tuozzolo, Q. Wu, T. Xin, Wencan Xu, A. Zaltsman

*Brookhaven National Laboratory, Upton, New York 11973-5000, USA*

To improve RHIC luminosity for heavy ion beam energies below 10 GeV/nucleon, the Low Energy RHIC electron Cooler (LEReC) is currently under commissioning at BNL. The Linac of LEReC is designed to deliver a 1.6 MeV to 2.6 MeV electron beam, with rms dp/p less than 5e-4. A 704 MHz superconducting radio frequency (SRF) booster cavity in this Linac provides up to 2.2 MeV accelerating voltage. With such a low energy and very demanding energy spread requirement, control of Higher Order Modes (HOMs) in the cavities becomes critical and needs to be carefully evaluated to ensure minimum impact on the beam. In this paper, we report the multiphysics design of the HOM damper for this cavity to meet the energy spread requirement, as well as experimental results of the cavity with and without the HOM damper.

## I. Introduction

To map the QCD phase diagram, especially to search for the QCD critical point using the Relativistic Heavy Ion Collider (RHIC), significant luminosity improvement at energies below 10 GeV/nucleon is required. This can be achieved with the help of an electron cooling upgrade called LEReC [1].

The electron accelerator for LEReC consists of a DC photoemission gun and a 704 MHz SRF booster cavity. The booster cavity for LEReC was converted from the SRF photocathode gun of the ERL project [2]. A one cell 704 MHz normal conducting cavity and a 3-cell third harmonic (2.1 GHz) normal conducting cavity will be added to de-chirp the energy spread and to compensate its non-linearity, respectively. The Linac of LEReC is designed to deliver 1.6 MeV to 2.6 MeV electron beam, with rms dp/p less than 5e-4.

The very low energy and small energy spread requirement makes it important to control the HOMs in these cavities, especially the 704 MHz SRF booster cavity. Starting from the analysis of the HOMs in the bare cavity, and the wake potential associated with these HOMs, we identified the dangerous modes. Based on RF, thermal and mechanical analyses, we developed an HOM damper design to suppress these modes. A conditioning box was designed and tested to identify and overcome possible multipacting barriers of the damper at room temperature. The booster cavity was cryogenically tested without and with the HOM damper, and the results are also reported in this paper.

―――――――――――――――――――
†binping@bnl.gov

## II. Bunch Structure, Cavity HOMs and Energy Spread

### A. LEReC Bunch Structure

The LEReC design is a non-magnetized cooling approach that uses electron bunches that match the ion beam velocity to cool each single ion bunch. The ion beam in RHIC to be cooled consists of 111 bunches plus 9 missing bunches for the abort gap, that are evenly distributed in the 3833.84 m circumference, with γ ranging from 4.1 to 6.1. It uses a 9 MHz RF system with a wide tuning range. In this paper, 9 MHz refers to the 120$^{th}$ harmonic of the RHIC revolution frequency, ranging from 9.104 MHz to 9.256 MHz in LEReC. The LEReC electron beam uses a macropulse structure, with electron macropulses aligned to the RHIC ion bunches. Each macropulse contains up to 30 flat-top electron bunches spaced by 1.42 ns (704 MHz), with a 9 MHz macropulse repetition rate, or roughly 40% duty factor, and kinetic energies between 1.6 MeV and 2.6 MeV. In this paper, 9 MHz refers to the 120$^{th}$ harmonic of the RHIC revolution frequency, ranging from 9.104 MHz to 9.256 MHz in LEReC. A mode of 1.6 MeV operation with full continuous wave (CW) operation at the 704 MHz frequency (no macro-bunch structure) is also being considered. Note there are electron bunches in the ion abort gap. Table 1 summaries the proposed operating modes. The flat-top electron bunch is introduced in detail in section III (D).

### B. Cavity HOMs and Energy Spread

The booster cavity, shown in bottom of Figure 1, was converted from the SRF photocathode gun (top of

Figure 1) of the ERL project [2]. In this figure, the left side is upstream and right side downstream (i.e. beam travels left to right). It is a 0.4 cell cavity operating at 2 K, with a maximum energy gain of 2.2 MeV. Key cavity parameters are listed in Table 2. In the ERL gun configuration, it had a room temperature HOM damper located on the 10 cm diameter downstream beam pipe, outside the cryomodule [3]. The HOM absorber consisted of 12 pieces Copper-Tungsten composition Elkonites 10W3 substrates that form a cylindrical, 16 cm inner diameter ferrite spool, with two 50.8×38.1×3.18 mm nickel-zinc C-48 ferrite tiles soldered on each substrate. The geometric configuration of this absorber is identical to the new one shown in Figure 2. This spool is placed over a 3.9 cm long, 10cm I.D. 92% aluminum oxide ceramic window which is brazed on to the stainless steel beam pipe [3].

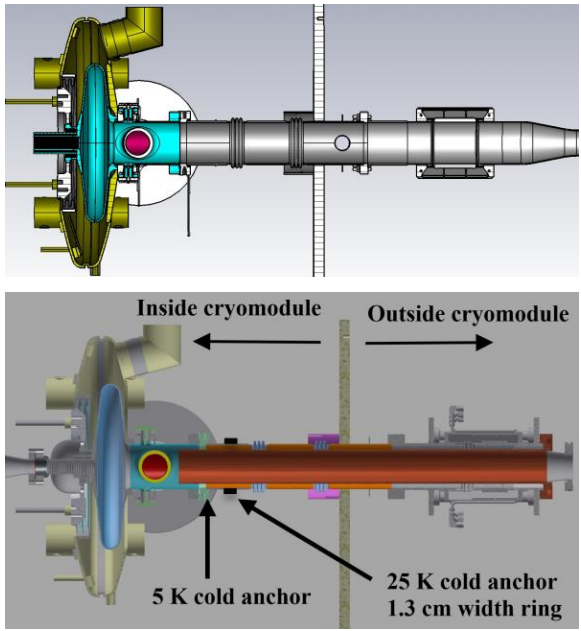

Figure 1. Top: ERL gun with old HOM assembly. Bottom: Booster cavity with new HOM assembly.

The 10 cm inner diameter beam pipe has a $TM_{01}$ cut-off frequency of 2.30 GHz. The cavity $TM_{010}$ fundamental mode frequency is 704 MHz. The first monopole HOM is the second harmonic $TM_{020}$ mode at 1.48 GHz, below the $TM_{01}$ beampipe cut-off frequency. This trapped $TM_{020}$ mode could therefore not be damped by the original ferrite damper that is far from the cavity. The previously measured loaded Q of this mode was ~165,000 at 2 K. However, note that for this measurement, the FPC port was connected to a network analyzer, which did not represent the actual operating configuration. The FPC design in this cavity provides good coupling to both fundamental mode and $TM_{020}$ mode. The cavity features two FPC ports, feeding from each side of the cavity with a broad band coaxial RF window separating the cavity vacuum from an air side narrow band doorknob transition to WR1500 waveguide (not shown in Figure 1), see reference [4] for more details. These two WR1500 waveguides, after some transition pieces and phase shifters, then connect via a waveguide tee, which then connects to a narrow band circulator that protects the 1 MW klystron. Thus the 1.48 GHz will be rejected by the narrow band components in the normal operating configuration. Simulation showed that in the normal operating configuration, the Q of the $TM_{020}$ mode would actually be $3.7\times10^7$. With a 50.8 Ω R/Q and a 0.12 V/pC loss factor for $TM_{020}$ mode, near resonant excitation of this HOM would induce well over 1 MV voltage fluctuation. There are still other dangerous modes below beam pipe cutoff which can readily increase the beam energy spread well past the specified limit.

Table 1. Proposed operating modes, final operating modes with dangerous HOM measured at 1.47834 GHz, and HOMs induced momentum spread in the worst-case scenario.

| Proposed operating modes | 1.6 MeV | 1.6 MeV CW | 2.0 MeV | 2.6 MeV |
|---|---|---|---|---|
| Bunch charge [pC] | 130 | 120.8 | 170 | 200 |
| Value of 9 MHz [MHz] | 9.104 | 9.104 | 9.187 | 9.256 |
| Bunches per macrobunch (9 MHz) | 30 | CW | 30 | 24-30 |
| Beam Current [mA] | 35.9 | 85.0 | 47.0 | 44.2-55.3 |
| **Final operating modes** | **1.60 MeV** | **1.60 MeV CW** | **1.92 MeV** | **2.60 MeV** |
| Dangerous HOM away from 9 MHz [MHz] | 3.41 | NA | 1.34 | >2.64 |
| dp/p from HOMs [$\pm10^{-4}$] | 2.7 | 3.6 | 2.5 | 1.7 |

Table 2. LEReC booster cavity parameter

| RF Frequency | 704.0 MHz |
|---|---|
| Active Length | 8.5 cm (0.4 cell) |
| Maximum Energy Gain | 2.2 MeV |
| R/Q (Acc. Def.) | 96.2 Ohm |
| Geometry Factor | 112.7 Ohm |
| Cavity Operating Temp. | 2 K |
| Power Coupler $Q_{ext}$ | $1.7\times10^5$ |
| Frequency Tuning Range | 1 MHz |
| Required RF Power | 122 kW |
| Installed RF Power | 130 kW |

Note that in simulations, the R/Q for each HOM, and thus its wake potential, is calculated with β=1.

This is not the case in actual operation as the beam energy at the cavity entrance is only 400keV. However, for the most critical mode $TM_{020}$, with β at 1.0, the R/Q is 50.8 Ω, with β at 0.9, it lowers to 37.7 Ω, and with β at 0.8, it is even lower, at 27.3 Ω. Using the R/Q value with β=1.0 is therefore conservative and safe.

## III. HOM Damper Design

### A. HOM Damper Choice and RF Design

Due to time and resource constraints, design and fabrication of a new cavity and cryomodule were not a practical option. Several ideas for reducing the impact of dangerous HOMs were investigated, including shifting the HOM frequencies via cavity tuning, shifting the RF spectrum of the beam and damping the trapped modes via FPC coupling. Other methods include new HOM damper design, which will be the topic of this section, as well as detuning the HOM frequency away from the nearest 9 MHz beam harmonic. This could be achieved by slightly changing the ion beam (and corresponding electron beam) Lorentz factor γ, thus changing the exact frequency of the 9 MHz spectral components, or by changing the cavity HOM frequency.

We first evaluate the possibility to shift the cavity HOM frequency via the main tuner. For this to be a viable solution, the required minimum frequency shift in the most dangerous HOM is about 0.7 MHz (detailed later in this paper). The main tuner can adjust the fundamental frequency over approximately 1 MHz, from 703.77 to 704.74 MHz. Over this range however, the $TM_{020}$ HOM changes only 0.1 MHz. The cavity also has two FPC ports, and is equipped with a dual-feed coupling system[5]. High power waveguide phase shifters were placed at each FPC to permit adjustment of external Q. Investigating the possible HOM frequency shift occurring over the available FPC $Q_{ext}$ range [5], both the fundamental frequency shift (4 kHz) and the $TM_{020}$ mode frequency shift (0.06 MHz) were far too small to be of practical use. In addition, one would obviously need to fix the fundamental frequency, as well as the FPC coupling strength to the designed value during operation.

Next, improved damping mechanisms were evaluated. Moving the ERL gun ferrite damper closer to the cavity was not an option. To couple sufficiently to the evanescent $TM_{020}$ mode in the beam pipe, it would introduce significant RF loss from the fundamental mode before any effective damping could be made to $TM_{020}$ mode. The FPC by itself is broadband and can couple to $TM_{020}$ mode. However, in the real system it becomes impractical to use the FPCs for damping. A narrowband doorknob transition is used between the main WR1500 waveguide and the coaxial FPC structure. Damping HOM on the air side, upstream of the doorknob structure is not viable, since the through attenuation of the doorknob at 1.48 GHz varies with frequency, ranging from 13 dB to 40 dB within ±5 MHz around 1.48 GHz, with power dissipated on the doorknob wall to be negligible. Due to other physical modifications to the cavity on the upstream and downstream ends, as well as to the FPC, the exact frequency of $TM_{020}$ would be impossible to predict, and thus the idea of damping on the air side after the doorknob structure was abandoned. Modifying the FPC structure between the doorknob and vacuum window to somehow introduce a high power diplexer to couple out the $TM_{020}$ mode were obviously impractical as well. Thus any idea of HOM damping via the FPCs was abandoned.

Ultimately, the only viable solution for damping the HOMs that allowed for sufficient coupling to the trapped modes was a coaxial beam pipe coupler scheme, shown in bottom of Figure 1. A new cylindrical ceramic RF window, and a ferrite absorber similar to the ERL SRF gun scheme, are used in this design. We note $L$ the length of the Cu tube from its tip to the electric short on the right side, and $d$ the distance between the centre of RF window and the electric short. To minimize the coupling to the fundamental $TM_{010}$ mode, $d$ should be around $(A/2+1/4)\lambda_{010}$, $L$ should be around $(B/2+1/2)\lambda_{010}$. To maximize the damping to $TM_{020}$, $d$ should be around $(C/2+1/2)\lambda_{020}$, $L$ should be around $(D/2+1/4)\lambda_{020}$. In these constraints, $A$, $B$, $C$, $D$ are zero or a positive integer, and $\lambda$ is the wave length of the mode. This combination of $d$ and $L$ minimizes the HOM damper coupling to the fundamental mode, and maximizes the coupling to the $TM_{020}$ mode. Note that since $\lambda_{010}$ is not exactly $2\lambda_{020}$, CST microwave studio simulation is needed for optimization. The location of the coaxial opening should be optimized with the consideration of a number of important factors: RF heating from the fundamental mode, coupling strength to the $TM_{020}$ mode, perturbation to the fundamental frequency and FPC coupling, as well as possible multipacting between coaxial opening and FPC couplers. In the final design, considering the actual dimensions of the cavity and beam pipe with cryomodule, with the HOM damper placed right outside the cryomodule, and trying to make the assembly compact, we choose $A=C=0$, $B=3$, $D=8$, and $d=113.0$ mm, $L=848.4$ mm. With such a design, the dissipation of the fundamental mode on the ferrite is ~10 W, and the quality factor of the 1.478 GHz mode is expected to be at 50,000.

### B. HOM Absorber Thermal and Mechanical Design

The HOM absorber, shown in Figure 2, is similar to the version for ERL gun, consisting of 12 pieces 25%

Copper – 75% Tungsten composition Mi-Tech CW75 substrate, which has the same make up as Elkonites 10W3 [6]. During inspection of the ERL gun absorber, cracks were found on the ferrite tiles. Instead of nickel-zinc C-48 ferrite, this damper uses TT2-111R ferrite tiles [6] having a closely matched thermal expansion coefficient with the Mi-Tech CW75. The ferrite tiles are attached to the substrate using Duralco™ 125 silver filled epoxy. Although not strictly necessary since the RF power dissipated on the absorber is small (With 10 W from the fundamental 704 MHz, and HOM power as calculated in section D), we kept the water cooling channel design from the ERL gun HOM damper.

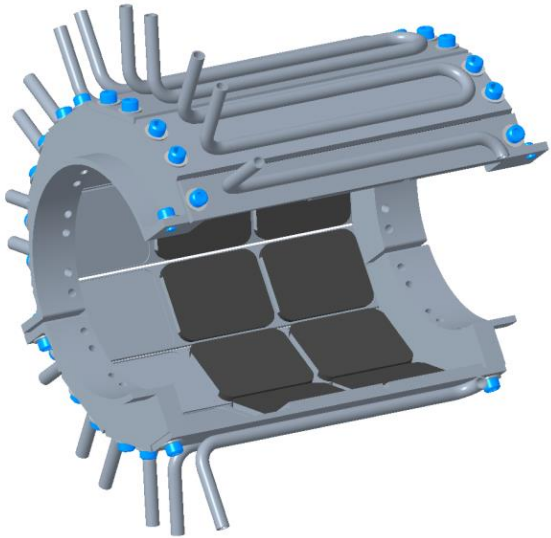

Figure 2. HOM absorber assembly section view with 12 pieces of substrates forming a cylinder, two ferrites on each substrate, and cooling channel on the back of each substrate.

Table 3. Thermal cycle test results

| Water cooling | No | Yes |
|---|---|---|
| Heater power [W] | 18 | 200 |
| Temp increase on ferrite [°C] | 70 | 10 |
| Temp gradient across ferrite [°C] | 14 | 120 |

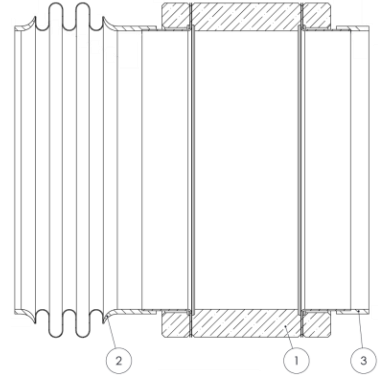

Figure 3. RF window assembly with: (1) sandwich ceramic window (2) stainless steel bellow and (3) 304 stainless steel adaptor.

During thermal cycle testing, a heater was mechanically bonded to the two ferrite tiles on one substrate. Different amounts of power were applied to the heater without and with water cooling. We measured the temperature increase (from room temperature) on the ferrite (the side away from the heater), as well as the temperature gradient across the ferrite. The results for the substrate with highest temperature increase are shown in Table 3. All ferrite tiles showed no crack after these tests. With 12 pieces, the HOM damper should be able to handle 216 Watt power without water cooling, and 2.4 kW with water cooling. Note that during these tests we did not try to push to its mechanical limit.

The dimensions of the RF window are close to the ERL gun HOM window. The ceramic braze assembly includes ceramic rings used to sandwich the stainless steel braze flanges, to balance the stress produced during brazing. A short bellows is included to minimize stress applied to the window during assembly and operation (Figure 3).

The copper tube (center conductor) of the coaxial coupler assembly is cantilevered from the electrically shorted end and is susceptible to mechanical vibrations. Simulations were performed to check for mechanical resonances, and a resonance at 60Hz was found. The tube was redesigned with a "taper" in the wall thickness, varying from 4 mm on electric short end to 1mm at the open end. This shifted the lowest mechanical resonance from 60 Hz to 90 Hz. Due to the difficulty in machining, the fabricated Cu tube is tapered with 4 steps instead of continuous tapering. Static deflection due to gravity, on the end close to the cavity is 0.04 mm. The cavity with HOM assembly is shown in Figure 1.

A water channel is designed on the electric short end to help stabilize the Cu tube temperature. Due to the long, narrow thermal path, fundamental mode induced RF heating on the Cu tube can induce a 25°C gradient from shorted to open end. In the thermal simulation of the booster cavity downstream end, we considered the RF induced heat, as well as the thermal radiation between the Cu tube and beam pipe. The RF induced heat at 2.2 MeV accelerating voltage is 4.6W on the tapered Cu tube, 2.2 W on the stainless steel beam pipe, and 0.2W on the AlMg gasket for NbTi flanges between the Nb and stainless steel beam pipe tubes, for a total of 7.1 W. With the water cooled electrical shorted end at 20°C, the open end Cu tip temperature increases to 45°C. To be conservative in estimating radiated heat, an emissivity of 0.1 is assumed for the tapered Cu tube (actual emissivity was measured to be < 0.025), and 55°C is assumed on the open end Cu tip. The open end of the tapered Cu tube faces the cavity downstream Nb beam pipe over a length of 4.7 cm, with the remainder facing the stainless steel beam pipe. In this case the thermal radiation from the tapered Cu tube is 7.5 W in total, with 2.0 W to Nb cavity, 0.8 W to the Nb beam pipe, and 4.7 W to stainless steel beam pipe. The cavity is immersed in 2 K liquid helium, with the Nb beam pipe and the FPC ports conductively cooled. The AlMg gasket is cooled with supercritical 5 K helium, with a cooling capacity of ~5 W. On the stainless-steel beam pipe, a 1.3 cm wide cylindrical thermal anchor strapped to the 25 K heat shield is added, 5 cm downstream from the 5 K anchor, to reduce the load to the 5 K intercept, shown in Figure 1. Heat load into the 2 K helium is 3.0 W, into the 5 K thermal intercept is 1.7 W (without the 25 K thermal anchor it would be 5.7 W), into the 25 K thermal anchor is 4.3 W, and into the water cooling outside the cryomodule is 5.6 W. With 3 W from the radiation, and possible beam halo heat load on the upstream side of the cavity, the 2 K cryostat loading, which was 18.3 W for the original ERL gun, is budgeted at 45 W. The maximum simulated temperature on the AlMg gasket with NbTi flanges is 6.56 K, on the Nb tube is 5.20 K, and on the stainless steel beam pipe is 305.86 K.

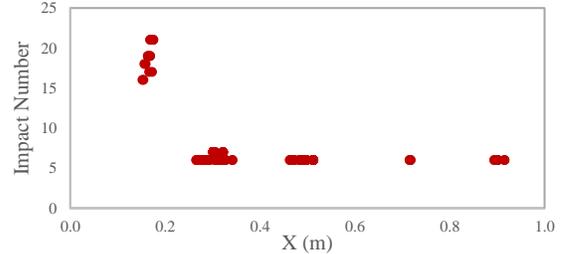

Figure 4. Multipacting simulation performed for the booster cavity HOM damper with TRACK3P. Electron impact energy (top), and number of impacts (bottom), along the HOM damper at accelerating voltage between 10 keV and 2 MeV. 0 m in X axis represents the cavity surface on equator, and 1 m the electric short of the Cu tube.

### C. HOM Damper Multipacting Simulation

Multipacting can cause the quality factor to drop and can limit the maximum cavity gradient. It is an electron avalanche effect due to resonant multiplication of secondary electrons [7]. It depends on the secondary electron yield of the cavity surface material and the cavity shape. It is normally easy to find in coaxial structure, and thus needs to be carefully evaluated in our coaxial HOM damper. We use the TRACK3P solver from the SLAC ACE3P suite of codes [8], scanning the accelerating voltage from 10 keV to 2 MeV with a 10 keV step size, and assumed seed electrons coming out of the whole section of the tapered Cu tube. As shown in Figure 4, with seed electrons starting from the cavity surface on equator, to the electric short of the Cu tube. Possible multipacting appeared at the FPC ports, and it was confined in the coaxial section of the FPC. This multipacting had been previously simulated and experimentally conditioned away. On the ceramic RF window, all particles die after 3 impacts. The number of electrons with energies above 200 eV is small, and all are confined on the FPC section, which indicates that multipacting should not be a critical issue for this cavity.

### D. Energy Spread with HOM Damper

In this section, we calculate the energy spread produced by the wake field in the booster cavity. We take intra-bunch (head to tail) energy spread caused by short range wake field, as well as inter-bunch (bunch to bunch) energy spread caused by long range wake field, into consideration.

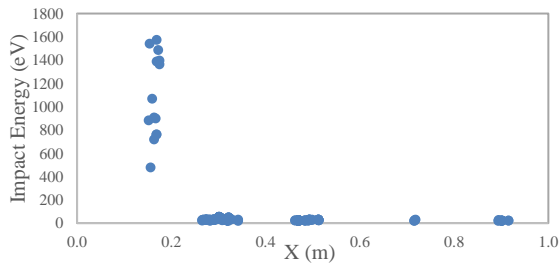

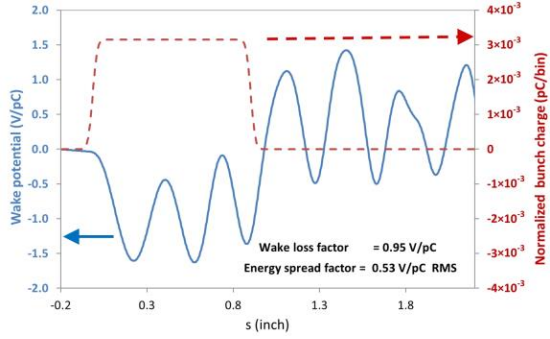

Figure 5. Short range wake field of the flat-top electron bunch, with wake loss factor at 0.95 V/pC and energy spread factor at 0.53 V/pC rms with 1.6 MeV beam energy.

The LEReC project uses a DC photoemission gun With multi-alkali ($CsK_2Sb$ or $NaK_2Sb$) cathode [9]. To get the desired 24 mm "flat-top" line density distribution, 32 Gaussian laser pulses, with 0.6 mm rms length and 0.75mm spacing, are stacked together[10]. In this case, one cannot simply use a 1 cm rms length Gaussian bunch for short range wake field simulation since a 0.6 mm bunch exhibits a frequency content much higher than the 1 cm bunch. A short range wake field simulation was done using CST Particle Studio$^{TM}$ with 0.6 mm rms Gaussian bunch at the speed of light. The result, as well as the charge distribution, were then calculated by "shift and stack" superposition, every 0.75 mm. The normalized results are shown in Figure 5. The wake loss factor is the integration of the product of wake potential and normalized bunch charge, and the energy spread factor is the rms deviation from the average energy loss. It is calculated by summing the weighted squares of the differences, and taking the square root of the sum. These two factors were then divided by $\beta^2$ for 1.6 MeV beam energy. The wake loss factor is 0.95 V/pC and energy spread factor is 0.53 V/pC rms. With 100 pC electron bunch, the energy spread intra-bunch is 53 V rms. Please note the $1/\beta^2$ factor is a simplified estimation based on reference [11], following the method in [11] leads to smaller wake loss and energy spread factors.

To calculate the inter-bunch energy spread from the long range wake field of the longitudinal modes, a straight-forward way is to use the result shown in Figure 5, "shift and stack" this result according to the beam pattern. This method is not used because it is very time consuming, and because one cannot apply the "worst case scenario" by artificially shifting each HOM frequency around, since the actual HOM frequency in the cavity might deviate from the RF simulation due to fabrication error, deformation during fabrication, operation, and tuning, etc. In this case Eigenmode simulation is first done using CST Microwave Studio$^{TM}$, with the simulation frequency ranging from the fundamental mode to the first longitudinal cut-off of the beam pipe. The single bunch wake potential is then constructed using the Eigenmode simulation results and is compared with the CST Particle Studio$^{TM}$ result. The multi bunch wake potential is calculated by using the "shift and stack" method on the single bunch wake potential, similar to the calculation of "flat-top" short range wake field.

For the 704 MHz SRF booster cavity, the downstream side beam pipe is 50 mm in radius, and a 40 mm radius tapered Cu tube for HOM damper is inserted into this beam pipe. The beam pipe is further taped to 30.2 mm radius for a third harmonic normal conducting RF cavity at 2.1 GHz next to it. The beam pipe cut-off frequency for 30.2 mm radius is 2.91 GHz for $TE_{11}$ mode and 3.81 GHz for $TM_{01}$ mode. We calculate the HOMs using CST Microwave Studio$^{TM}$ Eigen mode simulation with frequency up to 3.81 GHz. The result is then treated with "worst-case scenario" by artificially changing the resonance frequency of each HOM to align with multiples of 9 MHz, and for those modes that are close (±20 MHz) to the multiples of 704 MHz to the multiples of 704 MHz.

With the single bunch wake potential reconstructed from the Eigenmode simulation results using the method in [12] for point charge, we then use the "shift and stack" method introduced above to get the multi-bunch multi-train wake potential, with 30 continuous bunches with 704 MHz frequency in a 9 MHz train with ~40% duty factor. The results are consistent with the method proposed by Kim [13] since both methods are based on delta function structure. For the 704 MHz booster cavity, the multi-bunch multi-train saturates at 4.92 kV voltage fluctuation. With the $TM_{020}$ mode measured at 1.478 GHz 0.7 MHz away from the harmonic of the 9 MHz, the voltage fluctuation changes to 0.60 kV, corresponding to a maximum ±3.0e-4 dp/p peak to peak.

As mentioned in section III(A), it is practically difficult to tune the $TM_{020}$ mode frequency. Pushing the $TM_{020}$ mode away from the 9 MHz frequency (the repetition rate of the ion bunch, ranging from 9.104 MHz to 9.256 MHz in LEReC), is accomplished by carefully selecting the ion bunch energy, with the electron bunch velocity matching the ion bunch velocity. Table 1 shows the final operating modes, they are pulsed modes at 1.60 MHz, 1.92 MHz and 2.60 MHz, and CW mode at 1.60 MHz. The highest *dp/p* caused by HOMs in this cavity in the worst-case is ±3.6e-4 peak to peak with CW mode at 1.60 MHz.

Please note here we did not consider the beam loading effect of the fundamental mode at 704 MHz, even it follows the same calculation showed above. This effect will bring extra energy spread, and will be corrected by a combination of an additional 9 MHz cavity and RF power regulation for the booster cavity.

### E. Emittance growth with HOM damper

The inter-bunch (bunch to bunch) emittance growth of the electron beam from the long range wake potential of the transverse modes are calculated in a way similar to that of the inter-bunch energy spread. We use the same "worst case scenario" as mentioned in section III(D). Since the modes that are close to the multiples of the 704 MHz have low $R_T/Q$, none of these HOMs can accumulate voltage quickly within a train. In this case it is easy to understand that the highest $R_T/Q$ will give the most perturbation since we assumed all modes will be the multiples of 9 MHz. The most critical mode for vertical kick (aligned with FPC) is at 1.0057 GHz and the most critical mode for horizontal kick (perpendicular to FPC) is at 1.0049 GHz. The vertical mode is measured at around 1.0065 GHz at 2 K liquid helium bath temperature, the horizontal mode is not coupled to the FPC, thus cannot be measured. They are both $TM_{11}$ modes, and their polarizations are perpendicular to each other. The estimated maximum vertical kick is 5.2 kV, and for horizontal it is 1.43 kV for 0.5 mm displacement on each direction. Please note these two effect will not get stacked together, since these two resonances are 0.8 MHz away and they will not be the multiples of 9 MHz simultaneously. In this case $\Delta x'$ is estimated to be 1.8 mrad for vertical case and 0.51 mrad for horizontal case, and normalized $\Delta\varepsilon$ to be 7.7 mm×mrad for vertical and 2.2 mm×mrad for horizontal with 1 mm rms beam size in the cavity. This is a rough upper limit estimation. With the specification ε at 2.5 mm×mrad, the SRF booster cavity will contribute at most 3.1 times of the emittance for vertical case and 90% for horizontal case. One can always limit the displacement at a smaller number, i.e, 0.02 mm, so that the contribution will decrease to 12% for vertical case and 4% for horizontal case. This can be achieved by steering the beam right after the DC gun using dipoles to the electric center of the booster cavity. With the resonance frequencies of these two modes away from the harmonic of 9 MHz, the transverse emittance will reduce. In this case the displacement limitation can be relaxed accordingly.

### F. HOM power estimation

In this section, a method to estimate the HOM power generated in the booster cavity with HOM damper is introduced.

Bunch structure described in section II(B) is used for this analysis. The normalized beam spectrum $F_{norm}(\omega)$ is shown in Figure 6. To scale it to $I_{ave}=0.085A$ ave current (1.6 MeV CW case), an $I_{ave}*T=Q_T$ factor should be applied to the spectrum so that $F(\omega)=F_{norm}(\omega)*Q_T$, with $T$ the total time in the electron beam structure that is used for FFT, for LEReC we use the RHIC revolution time, and $Q_T$ the total charge of electron within $T$.

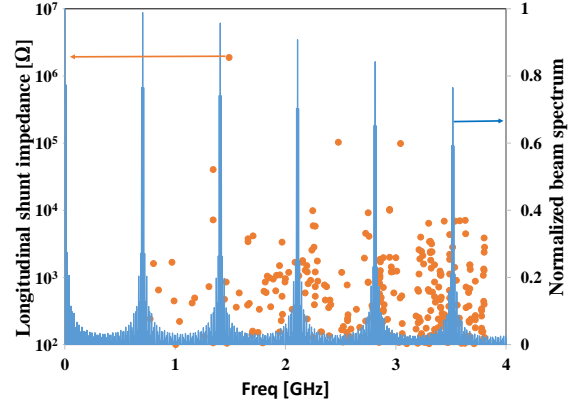

Figure 6. Normalized beam spectrum (blue curve) with shunt impedance of the longitudinal HOMs (red dots) in the booster cavity.

For each HOM, the real part of the shunt impedance over the frequency is calculated using the following formula:

$$R(\omega) = \frac{R_0}{1+Q_L^2(\frac{\omega_0}{\omega}-\frac{\omega}{\omega_0})^2}$$

With longitudinal shunt impedance to be $Q|V_z|^2/(2\omega U)$, shown as red dots in Figure 6, and transverse shunt impedance to be $Q|V_z(r_0)-V_z(0)|^2/(2\omega U)$ with displacement $r_0$ at 0.5 mm. Please note the definition of transverse shunt impedance is different from the typically used definition for beam stability simulation by a factor of $(c/(r_0\omega))^2$.

The power of each HOM is calculated using the following equation:

$$P_{HOM} = \sum_\omega I_{ave}^2 F_{norm}(\omega)^2 R(\omega)$$

With the worst-case assumption in section II(D), for the longitudinal modes, the $TM_{020}$ mode that is 0.7 MHz away from the multiples of 9 MHz produces 0.13 W power. For comparison, when it is the multiples of 9 MHz it is going to be 164 W. The 3.536 GHz mode produces 29.5 W, with 49% damped on the section close to the water cooling channel close to the electric short, and 29% on the ferrite. The 2.084 GHz mode produces 10.0 W, with 97% of power damped on the FPC. For the dipole modes, those at 1.0049 and 1.0057 GHz produce only 0.6 W in total in the worst case.

## IV. Test on HOM Damper

### A. Conditioning box design and test

A multi-purpose metal container, shown in Figure 7, is designed to bake, condition and store the tapered Cu tube with the RF window, so that this damper can be as clean as possible, and can be conditioned as much as possible to be mulipacting-free, to avoid any contamination to the SRF booster cavity, as well as to preserve a reasonably good emissivity by blocking water vaper and oxygen from contacting the tapered Cu tube. This container includes a stainless steel pipe that is identical in structure to the downstream beam pipe of the cavity, and another Cu tube that is connected to two 7/16 feedthroughs to introduce RF power to this condition box, shown on the left of Figure 7. Due to the design of this HOM damper, it rejects the 704 MHz $TM_{010}$ mode in the cavity. The distance $d$ (defined in section III(c)) cannot be altered. However, one can change the distance $L$ to give a better coupling to 704 MHz mode, this is done by the Cu tube inserted on the left side, which is slightly smaller than, and is inserted into the tapered Cu tube to provide reasonable RF transmission without physical contact between them to avoid any contamination or damage. The surface of the inserted Cu tube that faces the tapered Cu tube is grooved to suppress the multipacting in this overlapped section.

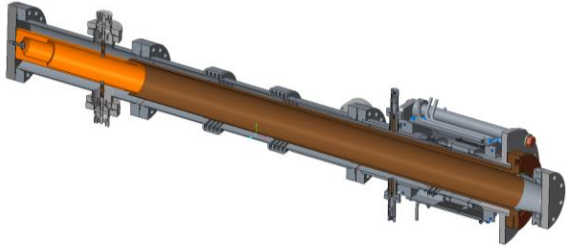

Figure 7. Multi-purpose metal container designed to bake, condition and store the tapered Cu tube with the RF window.

This assembly was first cleaned in class 100 clean room to particulate-free condition. It was then baked at 200°C with active pumping. A two port network analyzer was connected to this box, with one port connected to one of the 7/16 feedthrough on the left, and the other port connected to one of the pickup port that is on the left of the ferrite damper, as shown in Figure 7. The other 7/16 feedthrough was connected to a short with a coaxial cable, changing the length of this cable changes the coupling to this condition box. The length of this cable is determined so that the field inside this conditioning box can be maximized with a certain RF power. After that a 600 Watt amplifier was connected to the 7/16 feedthrough of this condition box, with the other 7/16 feedthrough remained the same (coaxial cable with a certain length, then a short). The goal of the high power test is to achieve an RF field equivalent, or higher than that in the SRF cavity with 2.2 MeV accelerating voltage. By applying as much as 420 Watts power at 704 MHz, the conditioning box reached an RF field equivalent to 3 MeV accelerating voltage in the SRF cavity. Some vacuum activities appeared with 300 Watts RF power, which is likely associated with multipacting effect. Since with this RF power, the corresponding accelerating voltage is ~2.7 MeV, well above 2.2 MeV, it is not a concern. The HOM damper assembly was conditioned to multipacting-free and was stored under vacuum in this conditioning box till it was assembled to the SRF booster cavity.

### B. Cavity test without and with HOM damper

The cavity was first tested without HOM damper. it reached 2.2 MeV accelerating voltage in CW mode, with 8~10 mRem/hr radiation, 7.0 W static load and 13.3 W dynamic load. It was then tested with HOM damper. Network analyzer measurement showed that quality factor of $TM_{020}$ mode is at 15,900, better than the simulated value at 50,000. During the high power test it reached 2.26 MeV in CW mode, with 8~18 mRem/hr radiation, 7.0 W static load and 18.3 W dynamic load. The temperature of NbTi flanges with AlMg gasket, as well as that of the downstream Nb tube that is conductively cooled, increased with increasing cavity gradient. The maximum temperature on the Nb tube was 6.6 K, 1.4 K higher than the thermal simulation. This is likely caused by the insufficient cooling time given to the 25 K thermal anchor added on the downstream (see Figure 1). This thermal anchor was above 54 K during the high power test. The test stopped at 2.26 MeV not because of any limitations by the cavity or HOM damper, but because of the concern that higher accelerating voltage might activate a field emitter and cause quality factor degradation, which was observed during the bare SRF booster cavity test at JLab.

## V. Conclusions

To meet the energy spread and emittance growth requirements of the LEReC project, an HOM damper was designed for the SRF booster cavity to damp the trapped $TM_{020}$ mode. Multi-physics simulations were performed to this design, which includes RF, thermal, mechanical and multipacting simulations. Calculations on energy spread caused by short-range and long range wake field, on emittance growth, and on HOM power estimation were done to ensure the effectiveness of this design. A multi-purpose condition box was designed to bake, condition and store the HOM damper. SRF booster cavity without and with HOM damper were tested cryogenically, results showed that this design meets the operational requirement at 2.2 MeV accelerating voltage, and meets the damping requirement of TM$_{020}$ mode.

## ACKNOWLEDGEMENT

The work is supported by by Brookhaven Science Associates, LLC under contract No. DE-AC02-98CH10886 with the US DOE. This research used the resources of the National Energy Research Scientific Computing Center (NERSC), which is supported by the US DOE under contract No. DE-AC02-05CH11231. The authors would like to thank M. Blaskiewicz for help with the wake potential calculations, C. Liu for help with the short range wake field simulation, R. Spitz for help on the thermal test of the ferrites, J. M. Brennan and C. Xu for help on the RF related issue, and K. Mernick for help on the cryogenic test.